\documentclass[12pt,a4paper]{article}
\usepackage{graphicx}
\usepackage{times}
\title{\bf In savvy pursuit of Local Group blue massive stars}
\author{Miriam Garcia$^{1,2}$, Artemio Herrero$^{1,2}$, Norberto Castro$^{1,2}$ and Luis Jos\'e Corral$^3$\\
\vspace{1cm}\\
\normalsize $^1$ Instituto de Astrof\'{\i}sica de Canarias, La Laguna, Spain \\ 
\normalsize $^2$ Universidad de La Laguna, La Laguna, Spain \\
\normalsize $^3$ Instituto de Astronom\'{\i}a y Meteorolog\'{\i}a, Universidad de Guadalajara, M\'exico \\}
\date{\mbox{}}
\begin{document}
\maketitle
\pagestyle{empty}
%
%
\def\bull{\vrule height .9ex width .8ex depth -.1ex}
\makeatletter
\def\ps@plain{\let\@mkboth\gobbletwo
\def\@oddhead{}\def\@oddfoot{\hfil\tiny\bull\quad
``The multi-wavelength view of hot, massive stars''; 39$^{\rm th}$ Li\`ege Int.\ Astroph.\ Coll., 12-16 July 2010 \quad\bull}%
\def\@evenhead{}\let\@evenfoot\@oddfoot}
\makeatother
%
%
\def\beginrefer{\section*{References}%
\begin{quotation}\mbox{}\par}
\def\refer#1\par{{\setlength{\parindent}{-\leftmargin}\indent#1\par}}
\def\endrefer{\end{quotation}}
%
%
{\noindent\small{\bf Abstract:} 

The important role of metallicity on massive star evolution
and the combination of multi-object spectrographs and 10m class telescopes,
have lead to numerous systematic studies of massive stars in Local Group galaxies.
While color based quests of blue massive stars are relatively successful,
they must be confirmed with spectroscopy and usually
lead to lists dominated by B-type modest-mass stars.
We have developed a friends of friends code to find OB associations
in Local Group galaxies, presented in Garcia et al. (2009).
One of the key points of the method is the photometric criterion
to choose candidate OB stars from the reddening-free Q parameter,
that could be easily extended to include from GALEX to near-IR photometry.
While not a new idea, one of our code's strong advantages
is the automatic determination of evolutionary masses for the members,
enabling a quick and more insightful choice of candidates for spectroscopy,
and the identification of potential advanced evolutionary stages.
We present our work on the very metal-poor irregular IC~1613
(Garcia et al. 2010a). The association properties are not only a powerful
aid towards finding the most interesting candidate massive stars,
but also reveal the galaxy's structure and 
recent star formation history.
}
%
%
\section{Introduction}

Massive stars are chief agents of galactic-scale feedback,
shaping the ISM with mighty winds throughout their evolution,
and utterly disrupting their environment in the supernova
explosion that puts their lives to an end.

Metallicity (Z) has a crucial two-fold role on blue massive star (BMS) evolution:
(i) through the equations of stellar structure, and
(ii) through the stellar wind (since $\dot M \propto Z^{0.8}$, Mokiem et al. 2007).
BMSs must be therefore studied systematically as a function of metallicity.

In particular, we are in dire need to understand what happens at metallicities smaller than
the Small Magellanic Cloud's, to make the connection
with the early Universe first stars.
The irregular galaxy IC~1613 is ideal
to study the very-poor metallicity regime (Garcia et al. \ 2010b),
located 714 Kpc away (Dolphin et al.\ 2001)
with metallicity 0.04 to 0.2 Z$_{\odot}$ (Talent \ 1980, Davidson \& Kinman \ 1982, Dodorico \& Dopita\ 1983,
Peimbert et al.\ 1988, Kingsburgh \& Barlow \ 1995, Tautvai{\v s}ien{\.e} et al.\ 2007, Herrero et al.\ 2010).
However, the required exposure times for quantitative spectroscopy of stars at this distance are long.
It is therefore critical to build a good BMS candidate list.

\section{What is AUTOPOP?}

AUTOPOP is a highly customizable, modular code written in IDL.
A detailed discussion of its use and the underlying theoretical
concepts are given in Garcia et al. (2009).

Provided with a photometric catalogue of stars and user-defined
target selection criteria for BMS, and isochrones,
it automatically finds OB associations and analyzes their color-magnitude diagram.
The final product is a list of candidate blue massive stars
and their masses.

Here we present a brief description of the code
and its application to IC~1613.
For more details we refer the reader to Garcia et al. (2009, 2010a).

\subsection{Input material} 

The code must be fed:
(i) a photometric catalog,
(ii) user-defined criteria for target selection and extinction correction, and
(iii) isochrones for age and mass determination.

The only requirement for the input catalog is to have coordinates,
and the colors required for target selection/extinction correction.
Since the code is modular the format is free,
although in this case the user must also provide the reading routine.

\begin{figure}[t]
\centering
\includegraphics[width=8cm]{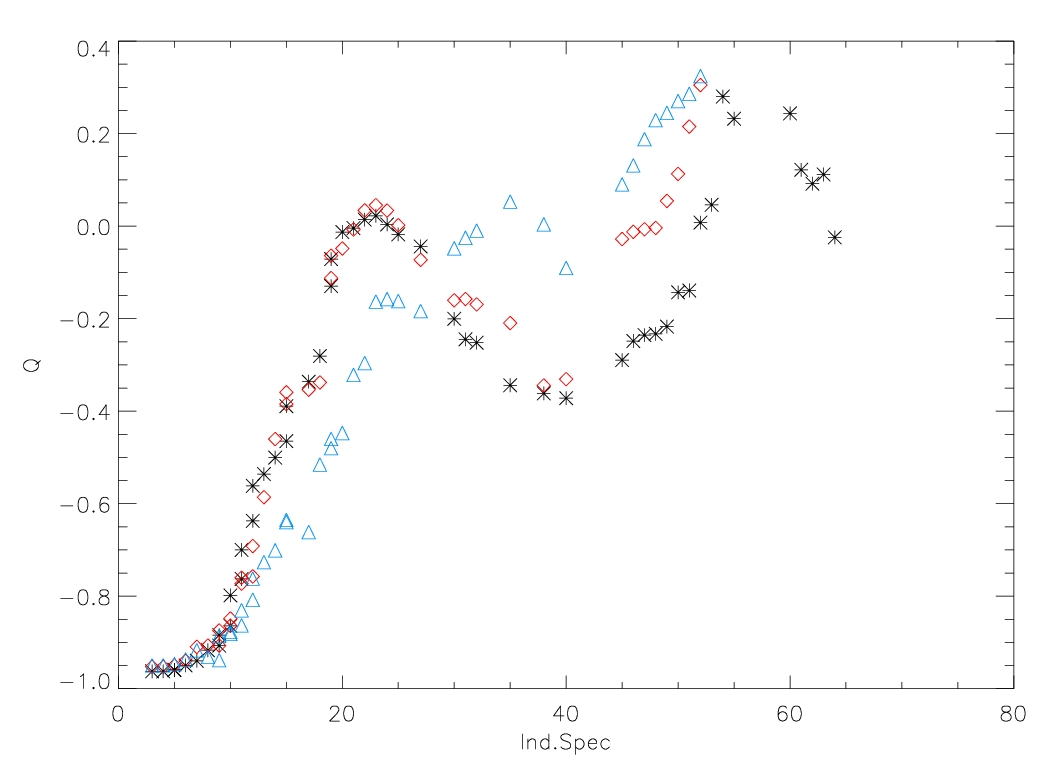}
\caption{
{\small
The variation of the reddening-free Q color with spectral type from O3 
(Ind.Spec=3) to M0 (Ind.Spec=60) types.
Colors were taken from Fitzgerald (1970).
Different symbols represent different luminosity classes.
For a given luminosity class, the relation between Q and spectral type
is biunivocal up to Ind.Spec=20 (i.e., A0 types).
}
\label{fig_1}}
\end{figure}

At the moment, the target selection criteria is based on the
reddening-free Q parameter
\begin{equation}
Q = (U-B) - 0.72 \times (B-V)
\end{equation}
which increases monotonically towards
later spectral types in the interval $\rm Q \in [-1.0,-0.4] $, corresponding to
O3-A0 types (see Fig.\,\ref{fig_1}).
In Fig.~5 of Garcia et al. (2009) we showed that most known OB stars in IC~1613
have $\rm Q \leq -0.4$.

However, following its versatility philosophy,
the code allows to easily modify the target selection criteria.
For instance, instead of using only optical photometry,
the code could be changed to allow a linear combination
of more colors including GALEX and 2MASS data.
Alternatively, the selection criteria could be based on 
fits to the spectral energy distribution 
made, for instance, with CHORIZOS (Ma\'{\i}z-Apell\'aniz\ 2004).
The latter method has been applied to M31 by Kang et al.\ (2009).

Finally, the user is free to use the isochrones of his/her choice,
with the only requirement that it is stored in an IDL structure
storing all the fields used by AUTOPOP.

\subsection{Code core}

The code consists of two main modules: 
automatic finding of geometrical groups in the sky,
which we shall call OB associations,
and the isochrone analysis of such associations.

The search of groups in the sky follows the Friends-of-Friends philosophy
or Path-Linkage criterion, after Battinnelli (1991):
two stars belong to the same
association if they dist less than a given search distance ($D_S$).
Given a target, the program looks for other stars within the search 
distance and registers them as members of the same association. 
The search is repeated for all the new members until no star is found within $D_S$~
from any of the peripheral stars.
Besides the list of points with spatial coordinates,
the code must also be provided with
the search distance and the  minimum number of members 
for a group to be considered so.
These parameters are specific for different galaxies (or input catalogs),
as we found they may be strongly dependent on the instrumental set-up (Garcia et al. \ 2010a).

Because the search of groups is run on a BMS candidate list,
the resulting geometrical groups are considered OB associations.
One may argue that there are systematics on this kind of codes (see Bastian et al.\ 2007)
and that OB associations are actually hierarchical groups,
a subset of the continuum of star formation only constrained by definition (see Elmegreen \& Efremov \ 1998).
This has been thoroughly discussed in Garcia et al. (2009,2010a) and shall not be repeated here,
but the conclusion is that physical or not,
these groups are helpful in sorting out
the content of blue massive stars in a galaxy.

The photometric analysis to derive masses and ages is made
in the Q \textit{vs} V$_{dered}$~ parameter space,
V$_{dered}$~ being the observed stellar magnitude corrected from extinction.
Such correction can be made thanks to the relation of Q with spectral type
(therefore intrinsic color) for early-type stars.
Massey et al. (2000) parametrized this relation for low-metallicity stars:
\begin{equation}
(B-V)_{0} = -0.005 + 0.317 \times Q
\end{equation}
using Kurucz' ATLAS9 Z=0.08Z$_{\odot}$ models
(no such relationship exists for metallicity as low as
IC~1613's).
This equation will not hold
if the reddening law deviates from the standard behaviour,
but in this case a traditional correction in the B-V \textit{vs} V diagram
will not be valid either.

Even though photometric errors are larger in the Q \textit{vs} V$_{dered}$~ parameter space
compared to the traditional B-V \textit{vs} V diagram because more colors are involved,
we preferred the former because isochrones are more spread,
and extinction is individually corrected towards each target.
Finally, the code for automatic isochrone analysis basically 
finds which isochrones enclose the star
and derives the evolutionary stellar properties from interpolation
of the closest points of those isochrones.
 
\section{Products from IC~1613 analysis}

In short, AUTOPOP ingests a photometric catalog
and returns a list of groups of likely OB stars, their ages, and masses for the members.
The most direct application is a ready-to-use list 
for spectroscopy of the most massive members
of a given galaxy.
The output tables for OB association members and properties 
are VO friendly, hence can be easily cross-matched with 
lists of X-ray sources or other public catalogs, 
increasing the chances to find interesting targets.

We have applied AUTOPOP to the analysis of the BMSs population
of stars in IC~1613 (Garcia et al.\ 2009, 2010a).
The input catalog of stars was built from observations of IC~1613
with the 2.5m Isaac Newton Telescope (INT) using the 
Wide Field Camera (WFC).
The field of view ($\rm 34 ' \times 34 '$)
covers the whole galaxy with a resolution of 0.33$"$/pixel.
The faint limit is V=25, and the 
maximum V-magnitude error for a V=23 star is 0.1mag.
More details are provided in Garcia et al. (2009).

\begin{figure}[t]
\begin{minipage}{8cm}
\centering
\includegraphics[width=8cm]{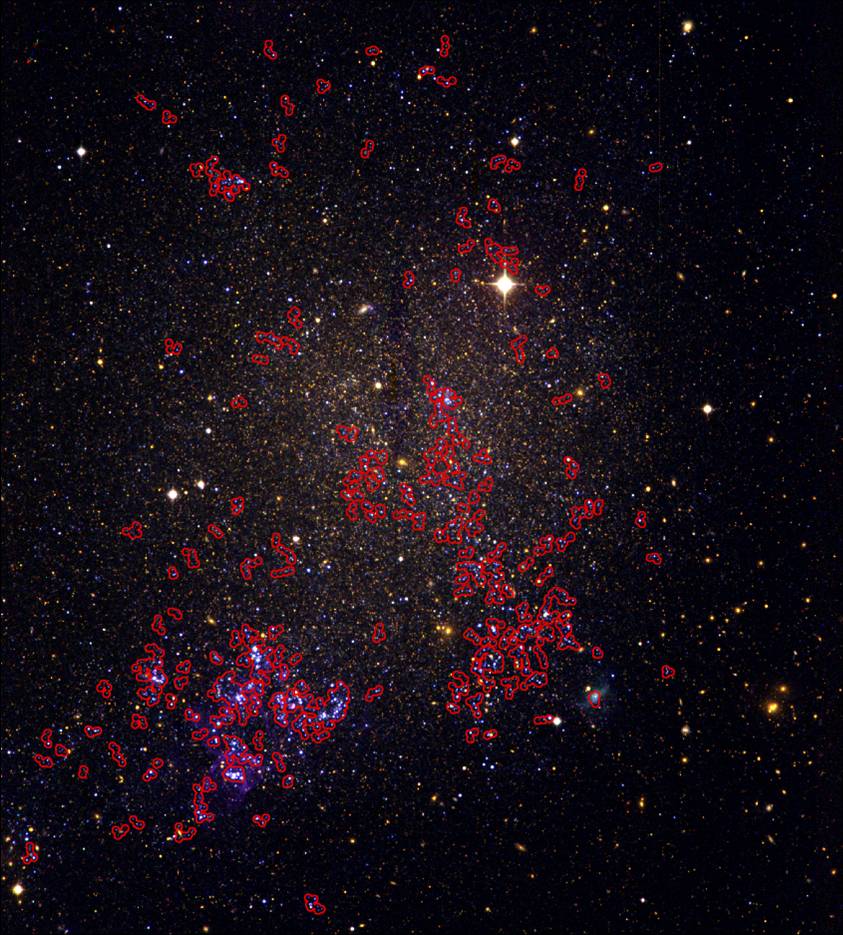}
\caption{
{\small
INT-WFC image of the optically brightest center of IC~1613.
North is left and East is down.
The RGB composition was made with the U- (blue), V- (green) and R-bands (red).
The positions of the associations found in this work are marked in red.
}
\label{fig_2}}
\end{minipage}
\hfill
\begin{minipage}{8cm}
\centering
\includegraphics[width=8cm]{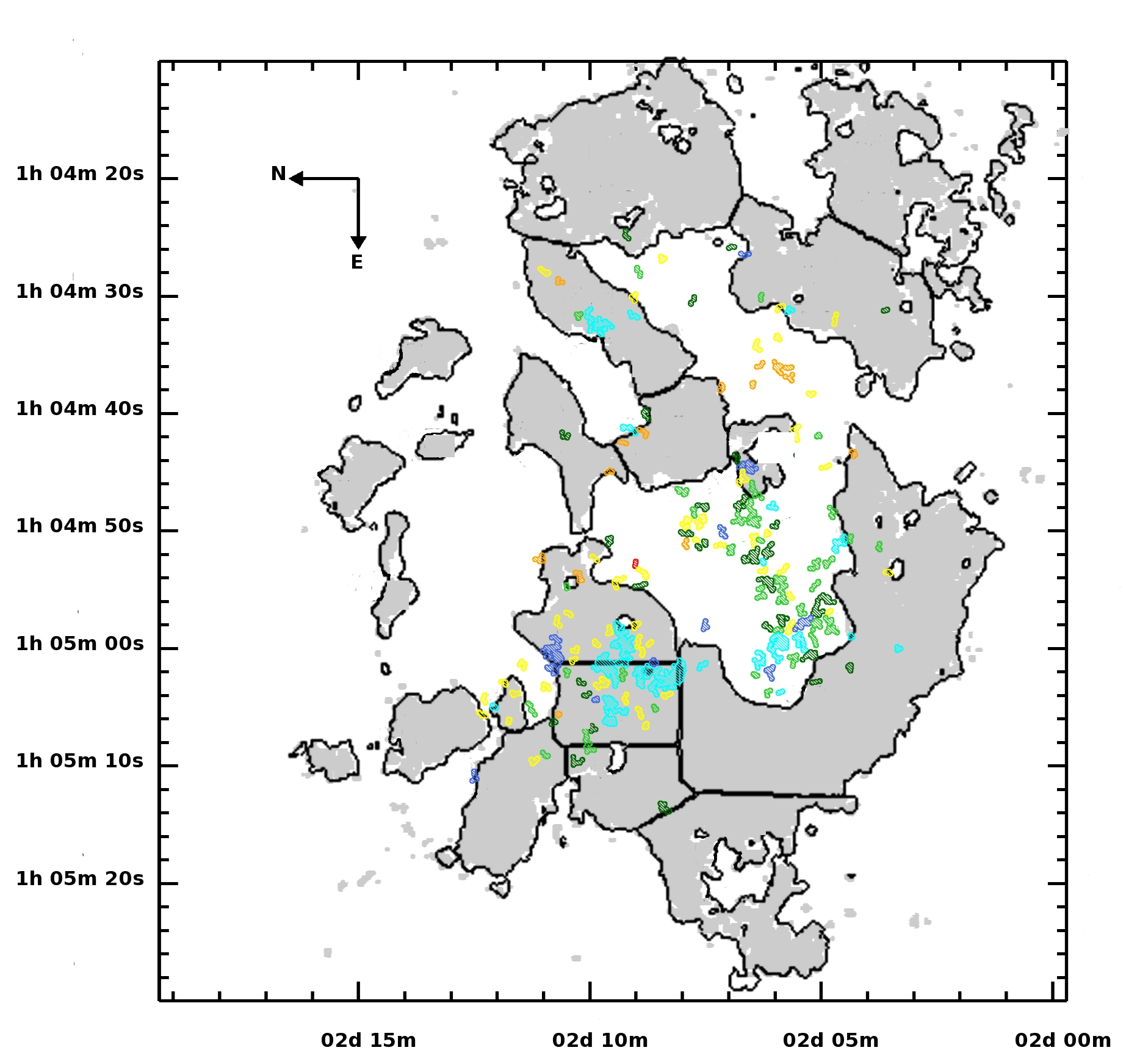}
\caption{
{\small
Relative distribution of neutral hydrogen and OB associations.
The grey figures represent the distribution of HI in IC~1613
(adapted from Fig.~1 of Silich et al.\ 2006). Contours
indicate the position of the associations, with 
colour indicating their ages (see text).
}
\label{fig_3}}
\end{minipage}
\end{figure}

For the study of the association populations
we used Lejeune \& Schaerer (2001)'s basic grid isochrones of Z=0.004 (0.2Z$_{\odot}$),
who recalculated Charbonnel et al. (1993)'s stellar tracks
with improved relations of photometric colors and stellar parameters.

The resulting OB associations are shown in Fig.~\ref{fig_2}.
They group in the NE lobe of the galaxy, where IC~1613 displays
spectacular  giant HII shells.
We find association ages ranging from $\log age [yr]=$ 5.9 to 7.7, 
usually with some spread.
For each association, we provide its age as a colored contour in Fig.~\ref{fig_3}
(cyan $ \log age [yr]=5.9$,
 blue        $ =6.2$,
 dark green  $ =6.5$,
 light green $ =6.8$,
 yellow      $ =7.1$, 
 orange      $ =7.4$, 
 red         $ =7.7$,
 violet      $ =8.0$).
There seems to be a galactic-scale age gradient following the 
NE direction, with the youngest members located in the NE lobe.
This region, which also exhibits the largest age dispersion in the galaxy,
is located in a density enhancement of neutral hydrogen
(see Garcia et al.\ 2010a and Fig.~\ref{fig_3}).
This fact suggests that star formation has proceeded in this region
for an extended period of time,
and that sequential star formation may be at work.

\section{Summary and conclusions}

The spectroscopic studies of OB stars in the Local Volume
are now possible thanks to multi-object spectrographs on 10m-class telescopes.
Such studies are driven by the need to understand the impact
of metallicity on OB star evolution.
It is necessary to optimize the selection of targets, since these
observations are very expensive in observing time.
AUTOPOP provides an optimal way of obtaining potentially very interesting
candidates.

%
%
\section*{Acknowledgements}
Funded by Spanish MICINN under CONSOLIDER-INGENIO 2010,
program grant CSD2006-00070,
and grants AYA2007-67456-C02-01
and  AYA2008-06166-C03-01.
%
%
\footnotesize
\beginrefer
\refer Bastian N., et al., 2007, MNRAS, 379, 1302 

\refer Battinelli P., 1991, A\&A, 244, 69 

\refer Charbonnel C., et al., 1993, A\&ASS, 101, 415 

\refer Davidson K., Kinman T.D., 1982, PASP, 94, 634 

\refer Dodorico S., Dopita M., 1983, in \textit{Supernova Remnants and their X-ray Emission}, 101, 517 

\refer Dolphin, A. E., et al., 2001, ApJ, 550, 554

\refer Elmegreen B.G.,  Efremov Y.N., 1998, arXiv:astro-ph/9801071 

\refer Fitzgerald M. P., 1970, A\&A, 4, 234

\refer Garcia M., Herrero A., Vicente, B., Castro, N., Corral, L. J., Rosenberg, A., Monelli, M., 2009, A\&A, 502, 1015

\refer Garcia M., et al., 2010a, A\&A, in press

\refer Garcia M., et al., 2010b, proceedings of \textit{Ultraviolet Universe 2010}

\refer Herrero A., Garcia M., et al., 2010, A\&A, 513, 70

\refer Kang Y., Bianchi L., Rey S.-C., 2009, ApJ, 703, 614

\refer Kingsburgh R.L., Barlow M.J., 1995, A\&A, 295, 171 

\refer Lejeune T., Schaerer D., 2001, A\&A, 366, 538 

\refer Ma\'{\i}z-Apell\'aniz J., 2004, PASP, 116, 859

\refer Massey P., Waterhouse E., DeGioia-Eastwood K., 2000, AJ, 119, 2214 

\refer Mokiem, M.~R., et al. 2007, A\&A, 473, 603

\refer Peimbert M., Bohigas J., Torres-Peimbert S., 1988, RMxAA, 16, 45 

\refer Talent, D.L., 1980, Ph.D.~Thesis

\refer Tautvai{\v s}ien{\.e} G. et al., 2007, AJ, 134, 2318 

\endrefer           
\end{document}